\documentclass{article}
\usepackage{icrctc07}

\title{Gamma-Ray Burst Follow-up Observations with STACEE During 2003-2007}

\shorttitle{STACEE GRB Observations}

\authors{A. Jarvis$^{a}$, J. Ball$^{a}$, J.E. Carson$^{a,1}$,
C.E. Covault$^{b}$, D.D. Driscoll$^{b}$, P. Fortin$^{c}$, D.M
Gingrich$^{d,e}$, D.S. Hanna$^{f}$, J. Kildea$^{f,2}$, T. Lindner$^{f,3}$,
R. Mukherjee$^{c}$, C. Mueller$^{f}$, R.A. Ong$^{a}$, K. Ragan$^{f}$,
D.A. Williams$^{g}$, J. Zweerink$^{a}$}

\afiliations{
$^{a}$Department of Physics and Astronomy, University of California at
Los Angeles, Los Angeles, CA, USA, 90095\\
$^{b}$Department of Physics, Case Western Reserve University,
10900 Euclid Ave., Cleveland, OH, USA, 44106\\
$^{c}$Department of Physics, Barnard College, Columbia University, New York, NY, USA,
10027\\
$^{d}$Centre for Particle Physics, University of Alberta, Edmonton,
AB T6G 2G7, Canada\\
$^{e}$TRIUMF, Vancouver, BC V6T 2A3, Canada\\
$^{f}$Department of Physics, McGill University, 3600 University
Street, Montreal, QC H3A 2T8, Canada\\
$^{g}$Santa Cruz Institute for Particle Physics, University of
California at Santa Cruz, 1156 High Street, Santa Cruz, CA, USA,
95064\\
$^{1}$Current Address: Stanford Linear Accelerator Center, MS 29, Menlo Park, CA 94025\\
$^{2}$Current Address: Fred Lawrence Whipple Observatory, Harvard-Smithsonian Center for Astrophysics, Amado, AZ 85645\\
$^{3}$Current Address: Department of Physics and Astronomy, University of British Columbia, Vancouver, BC V6T 1Z1, Canada\\
}

\email{ajarvis@physics.ucla.edu}

\abstract{The Solar Tower Atmospheric Cherenkov Effect Experiment
(STACEE) is an atmospheric Cherenkov telescope (ACT) that uses a large
mirror array to achieve a relatively low energy threshold.  For
sources with Crab-like spectra, at high elevations, the detector
response peaks near 100 GeV.  Gamma-ray burst (GRB) observations have
been a high priority for the STACEE collaboration since the inception
of the experiment.  We present the results of 20 GRB follow-up
observations at times ranging from 3 minutes to 15 hours after the
burst triggers.  Where redshift measurements are available, we place
constraints on the intrinsic high-energy spectra of the bursts.}

\begin{document}
\maketitle

\section{Introduction}

The Solar Tower Atmospheric Cherenkov Effect Experiment (STACEE) is a
showerfront-sampling Cherenkov telescope sensitive to gamma rays above
100 GeV. It is located at the National Solar Thermal Test Facility
(NSTTF) at Sandia National Laboratories outside Albuquerque, New
Mexico, USA. The NSTTF is located at 34.96$^{\circ}$N,
106.51$^{\circ}$W and is 1700 m above sea level. The facility has 220
heliostat mirrors designed to track the sun across the sky, each with
37 m$^{2}$ area. STACEE uses 64 of these heliostats to collect
Cherenkov light produced by air showers.

\begin{figure}[t]
\begin{minipage}{0.5\textwidth}
\begin{center}
  \includegraphics[width=.9\textwidth]{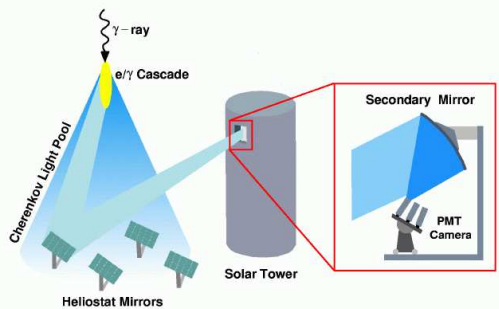}
\end{center}
\end{minipage}
\hfill
\begin{minipage}{0.45\textwidth}
  \caption{Conceptual drawing of STACEE.}
  \label{concept}
\vskip 1in
\end{minipage}
\end{figure}

STACEE employs five secondary mirrors on the solar tower to focus the
Cherenkov light onto photomultiplier tube (PMT) cameras, as shown in
Figure \ref{concept}. The light from each heliostat is detected by a
separate PMT and the waveform of the PMT signal is recorded by a flash
ADC. A programmable digital delay and trigger system\cite{IEEENSS2000}
selects showers for acquisition while eliminating most random
coincidences of night sky background photons. Under good observing
conditions, STACEE operates with a threshold around 5 photoelectrons
per channel. A detailed description of the instrument can be found in
D.M. Gingrich et al.\cite{Gingrich05}.

The large mirror area of the STACEE detector leads to an energy
threshold lower than those attainable by most single-dish imaging
telescopes or water-Cherenkov telescopes. The energy threshold -
defined as the energy at which the trigger rate peaks - is determined
by the spectrum of the source and the effective area of the detector
at the target position. For targets above 60$^{\circ}$ in elevation
with power-law spectral indices between 2 and 3, the energy threshold
is typically between 150 and 200 GeV. For targets near zenith, STACEE
has significant effective area at energies as low as 50 GeV.

A low energy threshold opens up the possibility of detecting more
distant sources\cite{Primack05}. Collisions of gamma rays with
extragalactic background light (EBL) photons produce electron-positron
pairs, attenuating the gamma-ray flux from distant sources. The
extinction becomes more severe with increasing energy, producing an
energy-dependent horizon for gamma-ray observations.  Thus, a low
energy threshold is advantageous when attempting to characterize the
high-energy emission of GRBs, for which the mean measured redshift
(for Swift bursts) is 2.7\cite{Le07}.

STACEE observations are typically taken in pairs of on-source and
off-source runs.  The off-source runs serve as measurements of the
background event rate produced by cosmic-ray showers.  Under normal
conditions, the cosmic-ray trigger rate is $\sim$5 Hz.  Background
rejection techniques have been explored and the most effective
technique is described
elsewhere\cite{Kildea05,Kildea05_2,Lindner06}. After background
rejection cuts, STACEE typically obtains a 5$\sigma$ detection of the
Crab with approximately 10 hours of on-source observations.  Under
good observing conditions, STACEE would obtain a 5$\sigma$ detection
in 30 minutes for a source with a spectrum equal to 4.5 times that of
the Crab.

\section{GRB Observing Strategy}

Observing gamma-ray bursts is a high priority for STACEE.  Burst
alerts from the GRB Coordinates Network (GCN)\cite{GCN} are monitored
with a computer program that alerts STACEE operators when a burst is
observable by updating a web page and initiating an audio alert.
Email alerts are also sent to a cellular phone and pager that are
carried by the telescope operators.  Upon receiving a burst alert,
operators immediately abort any observations in progress and retarget
the detector.  In addition to doing immediate GRB follow-up
observations, we search for afterglow emission from bursts discovered
within the previous 12 hours. EGRET detected GeV emission from
GRB940217, including an 18 GeV photon, for more than 90 minutes after
the start of the burst\cite{Hurley1994}.

The Swift GRB Explorer\cite{Gehrels04} launched in November of 2004,
is able to pinpoint burst directions on the unprecedented timescale of
about 20 seconds. In the summer of 2004, the 64 heliostats used by
STACEE were outfitted with new motors to increase their slewing
speed. The new motors have performed very well, allowing the detector
to be retargeted from any of the sources typically observed to most
observable GRB positions within 1 minute. In the fall of 2005, a
socket connection was established between the STACEE burst alert
system and the GCN, effectively eliminating the delays associated with
sending and receiving emails.  In principle, STACEE should be able to
begin observations of some bursts within 100 seconds of their Swift
triggers.

\section{STACEE GRB Observations}

Since the summer of 2003, follow-up observations have been made for 20
bursts with STACEE.  For two of those bursts, GRB 031220 and GRB
061222, the observing conditions were very poor and the data could not
be analyzed with a standard on-off comparison.  Analysis of the event
rates during the on-source runs for these bursts did not reveal any
significant spikes.  The remaining burst observations are described in
Table \ref{observations}.  Two of the most interesting observations
are discussed briefly in this section.

\begin{table*}[th]   
\begin{center}
\begin{tabular}{||c|c|c|c|c|c|c|c||} \hline \hline

         &          &      & Time to &          &              & Energy & 95\% CL Upper \\
Burst ID & Redshift & Ref. & Target  & Livetime & Significance & Threshold & Flux Limit \\
         &          &      & (min)   & (min)    &              & (GeV) & (ergs cm$^{-2}$ s$^{-1}$) \\
                             \hline \hline

 040422 &       & & 95.3 & 28 & -0.77 & 480 & $3.5\times10^{-10}$ \\
                             \hline
 040916 &       & & 309.7 & 28 & 0.98 & 320 & $9.0\times10^{-10}$ \\
                             \hline
 041016 &       & & 142.0 & 4.4 & -1.3 & 310 & $2.4\times10^{-9}$ \\
                             \hline
 050209 &       & & 217.5 & 24 & 0.34 & 260 & $6.1\times10^{-10}$ \\
                             \hline
 050402 &       & & 4.0 & 18 & 0.05 & 470 & $8.3\times10^{-10}$ \\
                             \hline
 050408 & 1.236 & \cite{Berger05} & 642.5 & 17 & 0.28 & 530 & $1.3\times10^{-9}$ \\
                             \hline
 050412 &       & & 7.2 & 6.9 & 2.1 & 250 & $1.5\times10^{-9}$ \\
                             \hline
 050509B & 0.225 & \cite{Gehrels05} & 23.2 & 35 & 0.69 & 150 & $3.5\times10^{-10}$ \\
                             \hline
 050509A &       & & 476.4 & 16 & 0.26 & 520 & $9.0\times10^{-10}$ \\
                             \hline
 050607 & $<$ 5 & \cite{Rhoads05} & 3.2 & 22 & -1.4 & 160 & $2.1\times10^{-10}$ \\
                             \hline
 060121 &        & & 384.7 & 30 & -0.96 & 650 & $5.0\times10^{-10}$ \\
                             \hline
 060206 & 4.045 & \cite{Fynbo06} & 269.8 & 52 & 0.79 & 180 & $3.5\times10^{-10}$ \\
                             \hline
 060323 &       & & 880.5 & 18 & -0.26 & 240 & $4.0\times10^{-10}$ \\
                             \hline
 060526 & 3.21 & \cite{Berger06} & 685.6 & 40 & 1.38 & 250 & $5.0\times10^{-10}$ \\
                             \hline
 061028 &       & & 339.2 & 57 & -0.04 & pending & pending \\
                             \hline
 070223 &       & & 357.9 & 42 & 0.85 & pending & pending \\
                             \hline
 070418 &       & & 360.1 & 25 & -0.81 & pending & pending \\
                             \hline
 070419A &      & & 3.3   & 27 & -0.71 & pending & pending \\
                             \hline

\end{tabular}

\caption{\label{observations} Summary of STACEE GRB observations since
the summer of 2003.  The targeting times and livetimes take into
account time removed by data quality cuts.  The energy thresholds and
flux limits assume a source photon spectrum dN/dE $\sim$ E$^{-2.5}$
that is integrated between 100 GeV and 10 TeV for the flux limits.}

\end{center}
\end{table*}

GRB 050509B was the first short burst for which a host galaxy was
determined.  The host is a large elliptical galaxy at a redshift of
0.225 \cite{Gehrels05}.  Because of the low redshift of this burst,
EBL attenuation would not be expected to totally obscure this burst in
STACEE's energy band.  The optical depth would be below 1 for photons
below about 400 GeV in energy.

STACEE observations of GRB 050509B began 23 minutes after the initial
burst detection and consisted of two 28-minute on/off pairs.  After
data quality cuts, approximately 35 minutes of livetime remained,
split almost equally between the two on-source runs (which were
separated by 32 minutes).  The burst was at a very nearly optimal
position, reaching its transit, about 6 degrees south of zenith,
halfway through the first on-source run.  Simulations were used to
determine the effective area, as a function of energy, for STACEE
under the conditions of these observations.  Assuming a photon
spectrum of dN/dE $\sim$ E$^{-2.5}$, the event rate limit determined
from observations was then converted into a flux limit.  After
accounting for EBL attenuation, the STACEE flux limit translates to a
limit on the average isotropic equivalent luminosity during the period
of observations of $L_{95} < 9.1 \times 10^{46}$ ergs/s in STACEE's
energy range.

\begin{figure}[t]
\begin{minipage}{0.5\textwidth}
\begin{center}  
  \includegraphics[height=.3\textheight]{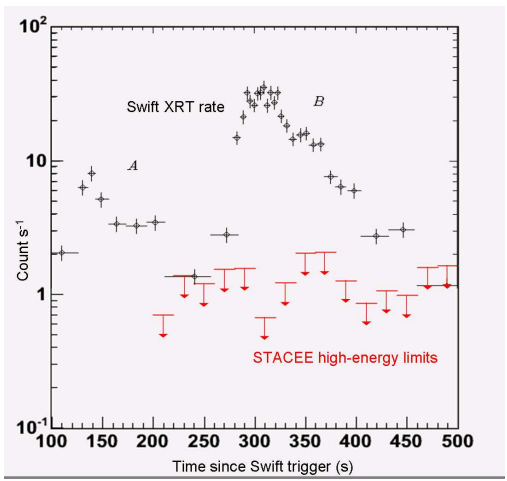}
\end{center}
\end{minipage}
\hfill
\begin{minipage}{0.45\textwidth}
  \caption{The Swift XRT light curve and STACEE high-energy rate limits for the early afterglow GRB 050607.}
  \label{050607}
\vskip 1in
\end{minipage}

\end{figure}

Our observation of GRB 050607 also stands out.  When observations
began, only 3 minutes and 13 seconds after the Swift burst trigger,
the burst was at a relatively high elevation of almost 62 degrees.
Approximately 5 minutes after the burst trigger a large flare in the
X-ray flux was detected by Swift's X-Ray Telescope (XRT).  Many
believe that such flares are evidence of late activity in the central
engine of the burst\cite{Nousek06} and it has been predicted that they
may be accompanied by flares of high-energy gamma rays\cite{Wang07}.
If that is the case the flares may provide additional hints about the
prompt emission.  No high-energy flaring was seen in STACEE
observations during the X-ray flare or in the subsequent twenty
minutes (Figure \ref{050607}). Unfortunately, the absence of a
redshift measurement for this burst prevents us from making any
stronger statements about the nature of these flares.

\section{GRB Followups in the Swift Era}

The Swift GRB Explorer has enabled ground-based observatories to
perform GRB followup observations with unprecedented response times.
STACEE has been able to begin observations of four bursts within 4
minutes of their initial Swift triggers. Swift also discovered that
the early X-ray afterglows of GRBs often contain large flares that are
thought to be evidence of late activity in the central engines of some
bursts.  Because of its high sensitivity, Swift is able to detect more
distant bursts than previous experiments.  The mean redshift of bursts
detected by Swift is 2.7 whereas the pre-Swift mean redshift was
1.5\cite{Le07}.  Because of their high redshifts, most bursts will be
obscured at high-energies due to EBL attenuation.  In addition,
redshift measurements have only been possible for a small fraction of
Swift bursts, making it difficult to translate flux limits into limits
on the intrinsic spectra of the bursts.  These factors magnify the
importance of obtaining fast, low-threshold followups for as many GRBs
as possible, as we have attempted to do with STACEE.  These factors
also heighten excitement about future experiments, such as GLAST, that
will bridge the energy gap between earlier satellite and ground based
telescopes.

\section{Acknowledgments}

This work was supported in part by the U.S. National Science
Foundation, the Natural Sciences and Engineering Research Council of
Canada, Fonds Quebecois de la Recherche sur la Nature et les
Technologies, the Research Corporation and the University of
California at Los Angeles.


\begin{thebibliography}{99}

\bibitem{IEEENSS2000}
J.-P. Martin and K. Ragan, Proc. IEEE Nuclear Science Symposium, vol. 8, pp 12-141-12-144, 2000.

\bibitem{Gingrich05}
D.M. Gingrich et al., IEEE Trans. Nucl. Sci., 52, 2977-2985, 2005.

\bibitem{Primack05}
J. R. Primack et al., Proc. AIP Conf. 745, 23, 2005.

\bibitem{Le07}
T. Le and C.D. Dermer, ApJ 661, 394-415, 2007.

\bibitem{Kildea05}
J. Kildea et al., Proc. of the 29th ICRC, 5, 135, 2005.

\bibitem{Kildea05_2}
J. Kildea et al., Proc. of the 29th ICRC, 4, 89, 2005.

\bibitem{Lindner06}
T. Lindner, PhD Thesis, McGill University, 2006.

\bibitem{GCN}
http://gcn.gsfc.nasa.gov/.

\bibitem{Hurley1994}
K. Hurley et al., Nature, 372, 652-654, 1994.

\bibitem{Gehrels04}
N. Gehrels et al., ApJ 611, 1005, 2004.

\bibitem{Gehrels05}
N. Gehrels et al., Nature 437, 851, 2005.

\bibitem{Nousek06}
J.A. Nousek et al., ApJ 642, 389-400, 2006.

\bibitem{Wang07}
X.Y. Wang, Z. Li and P. Meszaros, astro-ph/0702617, 2007.

\bibitem{Berger05}
E. Berger et al., GCN GRB Observation Report 3201, 2005.

\bibitem{Rhoads05}
J. Rhoads et al., GCN GRB Observation Report 3531, 2005.

\bibitem{Fynbo06}
U. Fynbo et al., GCN GRB Observation Report 4692, 2006.

\bibitem{Berger06}
E. Berger et al., GCN GRB Observation Report 5170, 2006.

\end{thebibliography}
\end{document}